\documentclass[12pt]{article}
\usepackage{amssymb}

\usepackage{a4}
\usepackage{amsmath}
\usepackage{amsfonts}
\usepackage{amsmath}
\usepackage{latexsym}
\usepackage{amsfonts}

\begin{document}

\noindent hep-th/0502123
   \hfill  February  2005 \\

\noindent \vskip3.3cm
\begin{center}

{\Large\bf The off-shell behaviour of propagators and the Goldstone
field in higher spin gauge theory on $AdS_{d+1}$ space}
\bigskip\bigskip\bigskip

{\large Ruben Manvelyan\footnote{On leave from Yerevan Physics
Institute} and Werner R\"uhl}
\medskip

{\small\it Department of Physics\\ Erwin Schr\"odinger Stra\ss e \\
Technical University of Kaiserslautern, Postfach 3049}\\
{\small\it 67653
Kaiserslautern, Germany}\\
\medskip
{\small\tt manvel,ruehl@physik.uni-kl.de}
\end{center}

\bigskip 
\begin{center}
{\sc Abstract}
\end{center}
A detailed analysis of the structure and gauge dependence of the
bulk-to-bulk propagators for the higher spin gauge fields in $AdS$
space is performed. The possible freedom in the construction of the
propagators is investigated and fixed by the correct boundary
behaviour and correspondence to the representation theory results
for the $AdS$ space isometry group. The classical origin of the
Goldstone mode and its connection with the gauge fixing procedure is
considered.

\newpage

\section{Introduction}
The $AdS_{4}/CFT_{3}$ correspondence of the critical $O(N)$ sigma
model and four dimensional higher spin gauge theory in anti de
Sitter space (HS(4)) proposed in \cite{Klebanov} increased the
interest in the old problems of quantization and interaction of the
higher spin gauge theories in $AdS$ space \cite{Frons, Vasiliev}.
This case of the general $AdS_{d+1}/CFT_{d}$ \cite{Maldacena}
correspondence is interesting also in view of the investigation of
gauge symmetry breaking and mass generation on the higher spin side
which are related to well explored properties of the corresponding
$CFT$ boundary theory at next-to-leading order in $\frac{1}{N}$.
This allows us to compare the result for the anomalous dimensions of
conserved currents of the critical $O(N)$ model with the the loop
corrections to the bulk-to-bulk propagator of the HS(4) following
from the possible special interaction between gauge and conformal
scalar fields in $AdS$. This was done in the last papers of the
authors \cite{Ruehl,RM3} where the mass correction for the HS(4)
field was evaluated. The analysis of the mass generation for the
HS(4) gauge fields and its comparison with the anomalous dimensions
of the conformal currents at large  $N$ limit  is important not only
for better understanding of the Higgs phenomenon for HS(4) theory
\cite{Por, Por1} but also for getting an answer to a more general
question: \emph{Does $AdS/CFT$ work correctly on the level of loop
diagrams in the general case and is it possible to use this
correspondence for real reconstruction of unknown local interacting
theories on the bulk from more or less well known conformal field
theories on the boundary side} \cite{R, T} ? For this purpose a
better understanding of the structure of the bulk-to-bulk
propagators of the higher spin gauge fields becomes very important
just as an investigation of the possible forms of local interactions
between HS gauge and scalar fields \cite{Leonhardt, MR}.

 In this article we perform a precise analysis of the gauge fixing
procedure in the Fronsdal theory of massless higher spin fields, and
analyze its quantization constructing the correct bulk-to-bulk
propagator of the higher spin theory. We describe also on the
classical level the origin of the Goldstone mode  responsible for
the quantum mass generation mechanism considered in the previous
paper \cite{RM3}.

\section{De Donder gauge and Goldstone mode}
We will use Euclidian $AdS_{d+1}$ with conformal flat metric,
curvature and covariant derivatives satisfying
\begin{eqnarray}
&&ds^{2}=g_{\mu \nu }(z)dz^{\mu }dz^{\nu
}=\frac{L^{2}}{(z^{0})^{2}}\delta _{\mu \nu }dz^{\mu }dz^{\nu
},\quad \sqrt{g}=\frac{L^{d+1}}{(z^{0})^{d+1}}\;,
\notag  \label{ads} \\
&&\left[ \nabla _{\mu },\,\nabla _{\nu }\right] V_{\lambda }^{\rho }=R_{\mu
\nu \lambda }^{\quad \,\,\sigma }V_{\sigma }^{\rho }-R_{\mu \nu \sigma
}^{\quad \,\,\rho }V_{\lambda }^{\sigma }\;,  \notag \\
&&R_{\mu \nu \lambda }^{\quad \,\,\rho
}=-\frac{1}{(z^{0})^{2}}\left( \delta _{\mu \lambda }\delta _{\nu
}^{\rho }-\delta _{\nu \lambda }\delta _{\mu }^{\rho }\right)
=-\frac{1}{L^{2}}\left( g_{\mu \lambda }(z)\delta _{\nu
}^{\rho }-g_{\nu \lambda }(z)\delta _{\mu }^{\rho }\right) \;,  \notag \\
&&R_{\mu \nu }=-\frac{d}{(z^{0})^{2}}\delta _{\mu \nu }=-\frac{d}{L^{2}}%
g_{\mu \nu }(z)\quad ,\quad R=-\frac{d(d+1)}{L^{2}}\;.  \notag
\end{eqnarray}%
For shortening the notation and calculation we contract all rank
$\ell$ symmetric tensors with the $\ell $-fold tensor product of a
vector $a^{\mu }$. In this notation Fronsdal's equation of motion
\cite{Frons} for the double traceless spin $\ell $ field is (from
now on we put $L=1$)
\begin{eqnarray}
&&\mathcal{F}(h^{(\ell )}(z;a))=\Box h^{(\ell )}(z;a)-(a\nabla )\nabla ^{\mu
}\frac{\partial }{\partial a^{\mu }}h^{(\ell )}+\frac{1}{2}(a\nabla
)^{2}\Box _{a}h^{(\ell )}(z;a)\quad   \notag  \label{fe} \\
&&-\left( \ell ^{2}+\ell (d-5)-2(d-2)\right) h^{(\ell )}-a^{2}\Box
_{a}h^{(\ell -2)}(z;a)=0,  \label{F} \\
&&\Box _{a}\Box _{a}h^{(\ell )}=0, \\&&
\Box=\nabla^{\mu}\nabla_{\mu}\quad,  \Box _{a}=g^{\mu \nu
}\frac{\partial ^{2}}{\partial a^{\mu }\partial a^{\nu }},\quad
(a\nabla)=a^{\mu}\nabla_{\mu} ,\quad
a^{2}=g_{\mu\nu}(z)a^{\mu}a^{\nu} .
\end{eqnarray}%
The basic property of this equation is higher spin gauge invariance with the
traceless parameter $\epsilon ^{(\ell -1)}(z;a),$
\begin{equation}
\delta h^{(\ell )}(z;a)=(a\nabla )\epsilon ^{(\ell -1)}(z;a),\quad \Box
_{a}\epsilon ^{(\ell -1)}(z;a)=0,\quad \delta \mathcal{F}(h^{(\ell
)}(z;a))=0.  \label{gi}
\end{equation}%
The equation (\ref{F}) can be simplified by gauge fixing. It is easy
to see that in the so called de Donder gauge
\begin{eqnarray}
&&\mathcal{D}^{(\ell -1)}(h^{(\ell )})=\nabla ^{\mu }\frac{\partial }{%
\partial a^{\mu }}h^{(\ell )}-\frac{1}{2}(a\nabla )\Box _{a}h^{(\ell )}=0,
\label{dD}
\end{eqnarray}
Fronsdal's equation simplifies to
\begin{eqnarray}
&&\mathcal{F}^{dD}(h^{(\ell )})=\Box h^{(\ell )}-\left( \ell
^{2}+\ell (d-5)-2(d-2)\right) h^{(\ell )}-a^{2}\Box _{a}h^{(\ell
-2)}=0.\quad \quad \label{physeq}
\end{eqnarray}
It was shown (see for example \cite{Mikh}) that in the de Donder
gauge the residual gauge symmetry leads to the tracelessness of the
\emph{on-shell} fields. So we can define our massless
\emph{physical} spin $\ell $ modes as traceless and transverse
symmetric tensor fields satisfying the equation (\ref{physeq})
\begin{eqnarray}
&&[\Box +\ell ]h^{(\ell )}=\Delta _{\ell }(\Delta _{\ell }-d)h^{(\ell )},
\label{ph1} \\
&&\Box _{a}h^{(\ell )}=\nabla ^{\mu }\frac{\partial } {\partial
a^{\mu }}
h^{(\ell )}=0,  \label{pd2} \\
&&\Delta _{\ell }=\ell +d-2.  \label{ph3}
\end{eqnarray}%
Note that equation (\ref{ph1}) for $\ell =0$ coincides with the equation for
the massless conformal coupled scalar only for $d=3$.

In a similar way we can describe the massive higher spin modes using
the same set of constraints on the general symmetric tensor field
$\phi^{(\ell)}(z,a)$ \cite{Waldron} but with the  independent
conformal weight  $\Delta$ (dimension) of the corresponding massive
(in means of $AdS$ field) representation of the $SO(d+1,1)$ isometry
group. This general representation with two independent quantum
numbers $[\Delta, \ell]$ under
the maximal compact subgroup goes, after imposing a shortening condition $%
\Delta=\Delta_{\ell}=\ell+d-2$, to the massless higher spin case (\ref{ph1}%
)-(\ref{ph3}) with the following decomposition \cite{Ruehl, Por,
Por1}
\begin{equation}  \label{dec}
\lim_{\Delta\rightarrow \ell+d-2}[\Delta,\ell]=[\ell+d-2,\ell]\oplus[%
\ell+d-1,\ell-1] .
\end{equation}
The additional massive representation $[\ell+d-1,\ell-1]$ is the
Goldstone field. Reading this decomposition from the opposite side,
we can interpret it as swallowing of the massive spin $\ell-1$
Goldstone field by the massless spin $\ell$ field with generation of
a mass for the latter one \cite{RM3}. For better understanding of
this phenomenon we need a more careful investigation of the gauge
invariant equation (\ref{fe}) in more general gauges.

First of all note that the gauge parameter $\epsilon^{(\ell-1)}$ is
a traceless rank $\ell-1$ tensor and therefore in any off-shell
consideration (quantization, propagator and perturbation theory) we
can use only gauge conditions with the same number of degrees of
freedom. The de-Donder gauge (\ref{dD}) is
just such a type of the gauge due to the tracelessness of the $\mathcal{D}%
^{(\ell-1)}(h^{(\ell)})$. Nevertheless for on-shell states we can
impose more restrictive gauges. Here we consider a one-parameter
family of gauge fixing conditions
\begin{equation}  \label{gengauge}
\mathcal{G}^{(\ell-1)}_{\alpha}(h^{(\ell)}) =\nabla^{\mu}\frac{\partial}{%
\partial a^{\mu}}h^{(\ell)}
-\frac{1}{\alpha}(a\nabla)\Box_{a}h^{(\ell)}=0
\end{equation}
This gauge condition coincides with the traceless de Donder gauge if $%
\alpha=2$ ($\Box_{a}\mathcal{G}^{(\ell-1)}_{2}=\Box_{a}\mathcal{D}%
^{(\ell-1)}=0$). Then we can write our double traceless field $%
h^{(\ell)}(z;a)$ as a sum of the two traceless spin $\ell$ and $\ell-2$
fields $\psi^{(\ell)}(z;a)$ and $\theta^{(\ell-2)}(z;a)$
\begin{eqnarray}
&& h^{(\ell)}(z;a)= \psi^{(\ell)}+\frac{a^{2}}{2\alpha_{0}}%
\theta^{(\ell-2)}(z;a)\quad,  \label{psi} \\
&&\Box_{a}h^{(\ell)}=\theta^{(\ell-2)}\quad,\quad
\Box_{a}\psi^{(\ell)}=\Box_{a}\theta^{(\ell-2)}=0 ,\\
&&\alpha_{0}=d+2\ell-3 .\label{a0}
\end{eqnarray}
In this parametrization Fronsdal's equation of motion with the gauge
condition (\ref{gengauge}) can be written in the form of the
following system of
equations for the two independent traceless fields $\psi^{(\ell)}$ and $%
\theta^{(\ell-2)}$
\begin{eqnarray}
&& \nabla^{\mu}\frac{\partial}{\partial a^{\mu}}\psi^{(\ell)}+\frac{a^{2}}{%
2\alpha_{0}} \nabla^{\mu}\frac{\partial}{\partial
a^{\mu}}\theta^{(\ell-2)}=
\frac{\alpha_{0}-\alpha}{\alpha\alpha_{0}}(a\nabla)\theta^{(\ell-2)}
,
\label{gg} \\
&&\left(\Box+\ell\right)\psi^{(\ell)}+\frac{\alpha-2}{2\alpha} \left[%
(a\nabla)^{2}\theta^{(\ell-2)}-a^{2}\frac{\alpha(\alpha_{0}-1)} {%
\alpha_{0}(\alpha-1)}\theta^{(\ell-2)}\right]=\Delta_{\ell}(\Delta_{\ell}-d)
\psi^{(\ell)} ,\qquad  \label{pem} \\
&&\left(\Box+\ell-2\right)\theta^{(\ell-2)}=\left[\Delta_{\theta}(\Delta_{%
\theta}-d)
+\frac{\alpha_{0}-\alpha}{\alpha-1}\right]\theta^{(\ell-2)} ,
\label{tem} \\
&&\Delta_{\ell}=d+\ell-2\quad,\quad \Delta_{\theta}=d+\ell-1 .
\label{d}
\end{eqnarray}

Now we are ready to discuss different gauge conditions. First of all
we see that the de Donder gauge ($\alpha=2$) leads to the complete
separation of the equations of motion for $\psi^{(\ell)}$ and
$\theta^{(\ell-2)}$ fields. On the other hand the gauge condition
(\ref{gg}) becomes just traceless for $\alpha=2$ and keeps on to
connect the divergence of $\psi^{(\ell)}$ and the traceless part of
the gradient of $\theta^{(\ell-2)}$
\begin{eqnarray}
  && \nabla^{\mu}\frac{\partial}{\partial a^{\mu}}\psi^{(\ell)}(z;a)
  =\frac{\alpha_{0}-2}{2\alpha_{0}}G^{(\ell-1)}(z;a) ,\label{gm} \\
  &&G^{(\ell-1)}(z;a)=(a\nabla)\theta^{(\ell-2)}(z;a)-\frac{a^{2}}
  {\alpha_{0}-2} \nabla^{\mu}\frac{\partial}{\partial
a^{\mu}}\theta^{(\ell-2)}(z;a) .
\end{eqnarray}
Here $G^{(\ell-1)}$ corresponds to the Goldstone representation.
Indeed using the equations of motion (\ref{pem}) and (\ref{tem})
with $\alpha=2$ one can derive that the $G^{(\ell-1)}$ field obeys
the following on-shell equation
\begin{equation}\label{gme}
    \left(\Box+\ell-1\right)G^{(\ell-1)}(z;a)=
    \Delta_{\theta}(\Delta_{\theta}-d)G^{(\ell-1)}(z;a)
\end{equation}
corresponding to the Goldstone representation
$[\Delta_{\theta}=\ell+d-1, \ell-1]$ arising in (\ref{dec}). This
mode can be gauged away on the classical level together with the
trace $\theta^{(\ell-2)}$ but only \emph{on-shell}. Therefore on the
quantum level this mode can arise in loop diagrams and will play the
crucial role in the mechanism of mass generation for the higher spin
gauge fields as it was shown in our previous paper \cite{RM3}.

Now we return to (\ref{gg})-(\ref{d}) and consider the next
interesting gauge $\alpha=d+2\ell-3$. This is a generalization for
the higher spin case of the so-called "Landau" gauge considered in
\cite{Freed} for the case of the graviton in $AdS_{d+1}$. But the
difference between the higher spin and graviton ($\ell=2$) cases is
essential. For the graviton we can apply this "Landau" gauge
\begin{equation}\label{lggrav}
    \nabla^{\mu}h_{\mu\nu}=\frac{1}{d+1}\partial_{\nu}h^{\mu}_{\mu}
\end{equation}
off-shell also because the trace is scalar here and this gauge fixes
the same number of degrees of freedom as the de Donder gauge. For
$\ell >2, \alpha\neq 2$ it is easy to see that condition (\ref{gg})
after taking the trace forces the trace components
$\theta^{(\ell-2)}$ of our double traceless field $h^{(\ell)}$ to be
transverse
\begin{eqnarray}
   &&\nabla^{\mu}\frac{\partial}{\partial
    a^{\mu}}\theta^{(\ell-2)}=0 ,\label{tt}\\
    && \nabla^{\mu}\frac{\partial}{\partial a^{\mu}}\psi^{(\ell)}
    =\frac{\alpha_{0}-\alpha}{\alpha\alpha_{0}}(a\nabla)\theta^{(\ell-2)} .\label{pt}
\end{eqnarray}
Moreover in the "Landau" gauge ($\alpha=\alpha_{0}$) the
$\psi^{(\ell)}$ component is also transverse but it's equation of
motion is  not diagonal like in the de Donder gauge. On the other
hand the equation of motion for the field $\theta^{(\ell-2)}$ is
simplified and we have for this field the realization of the
representation $[\Delta_{\theta}=\ell+d-1,\ell-2]$
\begin{equation}
    \left(\Box+\ell-2\right)\theta^{(\ell-2)}=\Delta_{\theta}(\Delta_{%
\theta}-d)\theta^{(\ell-2)} .
\end{equation}
So we see that only in the de Donder gauge we have a diagonal
equation of motion for the physical $\psi^{(\ell)}$ components but
this component is not transversal due to the presence of the
$[\ell+d-1, \ell-1]$ Goldstone mode $G^{(\ell-1)}$. This gauge is
most suitable for the quantization and construction of the
bulk-to-bulk propagator and for the investigation of the
$AdS_{4}/CFT_{3}$ correspondence in the case of the critical
conformal $O(N)$ boundary sigma model.
\section{Propagator}
Here we perform a precise analysis of the leading terms of the
bulk-to-bulk propagator obtained in \cite{LMR1} using boundary
integration of the bulk-to-boundary propagators constructed in
\cite{Dobrev} using representation theory. Here we construct and
analyse the general bitensorial ansatz satisfying the de Donder
gauge condition (\ref{dD}) and the equation of motion (\ref{physeq})
with the corresponding delta function on the right hand side. This
type of analysis can be simplified due to the following two
properties of the $AdS$ space propagators \cite{LMR1,LMR2}:
\begin{itemize}
    \item The propagator is a function of only one bilocal invariant
    variable,
    the geodesic distance
    \begin{equation}\label{zeta}
        \zeta(z_{1},z_{2})=\frac{(z^{0}_{1})^{2}+(z^{0}_{2})^{2}+(\vec{z}_{1}-\vec{z}_{2})^{2}}
        {2z^{0}_{1}z^{0}_{2}}=1+\frac{(z_{1}-z_{2})^{\mu}(z_{1}-z_{2})^{\nu}
        \delta_{\mu\nu}}{2z^{0}_{1}z^{0}_{2}}
    \end{equation}
     \item The tensorial structure of the bulk-to-bulk propagator can be
     explored using the following basis of the independent
     bitensors \cite{AllenJ}, \cite{AllenT}, \cite{Turyn}, \cite{LMR1}
     \begin{eqnarray}
       && I_{1}(a,c):=(a\partial)_{1}(c\partial)_{2}\zeta(z_{1},z_{2}) , \\
       && I_{2}(a,c):=(a\partial)_{1}\zeta(z_{1},z_{2})(c\partial)_{2}\zeta(z_{1},z_{2}),\\
       && I_{3}(a,c):=a^{2}_{1}I^{2}_{2c}+c^{2}_{2}I^{2}_{1a} , \\
       && I_{4}:=a^{2}_{1}c^{2}_{2} ,\\
       && I_{1a}:=(a\partial)_{1}\zeta(z_{1},z_{2})\quad ,
       \quad I_{2c}:=(c\partial)_{2}\zeta(z_{1},z_{2}) , \\
       &&(a\partial)_{1}=a^{\mu}\frac{\partial}{\partial
       z_{1}^{\mu}} ,\quad (c\partial)_{2}=c^{\mu}\frac{\partial}{\partial
       z_{2}^{\mu}} ,\\&& a^{2}_{1}=g_{\mu\nu}(z_{1})a^{\mu}a^{\nu} ,
       \quad c^{2}_{2}=g_{\mu\nu}(z_{2})c^{\mu} c^{\nu} .
     \end{eqnarray}
\end{itemize}

Using this basis we can start to construct the general ansatz. First
we introduce a special map from set $\{F_{k}(\zeta)\}^{\ell}_{k=0}$
of the $\ell+1$ functions on $\zeta$ to the space of $\ell\times
\ell$ bitensors
\begin{equation}\label{Psy}
    \Psi^{\ell}[F]=\sum^{\ell}_{k=0}I^{\ell-k}_{1}(a,c)I^{k}_{2}(a,c)F_{k}(\zeta)
    .
\end{equation}
The general expansion can be expressed then as
\begin{equation}\label{K}
    K^{\ell}(a,c)=\Psi^{\ell}[F]+\sum_{n,m;\, 0<2(n+m)<\ell}I_{3}^{n}I_{4}^{m}
    \Psi^{\ell-2(n+m)}[G^{(n,m)}] .
\end{equation}
We call all monomials in the above sum and the corresponding sets of
functions $\{G^{(n,m)}_{k}\}_{k=0}^{\ell-2(n+m)}$ the "trace terms".
The trace terms can be analyzed using the computer program
\cite{LMR2}. We select the first part and the first and second order
trace terms of (\ref{K})
\begin{eqnarray}
  && K^{\ell}(a,c)=\Psi^{\ell}[F]+I_{3}\Psi^{\ell-2}[G]+
  I_{4}\Psi^{\ell-2}[T]+I_{3}^{2}\Psi^{\ell-4}[H]+\dots \quad ,\label{Kk}\\
  && \{G^{(1,0)}_{k}\}^{\ell-2}_{k=0}=\{G_{k}\}^{\ell-2}_{k=0} ,
  \{G^{(0,1)}_{k}\}^{\ell-2}_{k=0}=\{T_{k}\}^{\ell-2}_{k=0} , \{G^{(2,0)}_{k}\}^{\ell-4}_{k=0}
  =\{H_{k}\}^{\ell-4}_{k=0},\label{sets}
\end{eqnarray}
and analyze the restrictions coming from the de Donder gauge fixing
and equation of motion. In other words we want to derive
differential and algebraical recursion relations for the
corresponding sets of functions (\ref{sets}) following from the
equations for the bulk-to-bulk propagator in the de Donder gauge.
First let's define the trace map
\begin{eqnarray}
    &&\Box_{a}\Psi^{\ell}[F]=I^{2}_{2c}\Psi^{\ell-2}
    [{Tr_{\ell}F}]+O(c^{2}_{2}) ,\label{trace}\\
&&(Tr_{\ell}F)_{k}=(\ell-k)(\ell-k-1)F_{k} +2(k+1)(\ell-k-1)\zeta
F_{k+1}\nonumber\\&&\quad\quad\quad\quad\quad+(k+2)(k+1)(\zeta^{2
}-1)F_{k+2} ,
\end{eqnarray}
the divergence and gradient maps
\begin{eqnarray}
  && \nabla^{\mu}\frac{\partial}{\partial a^{\mu}}\Psi^{\ell}[F]=
  I_{2c}\Psi^{\ell-1}[Div_{\ell}F]+O(c^{2}_{2}) , \label{div}\\
  && (Div_{\ell}F)_{k}=(\ell-k)\zeta
  F'_{k}+(k+1)(\zeta^{2}-1)F'_{k+1}\nonumber\\
  &&+(\ell-k)(\ell+d+k)F_{k}+(k+1)(\ell+d+k+1)\zeta F_{k+1} ,\\
  &&
  (a\nabla)\Psi^{\ell}[F]=I_{1a}\Psi^{\ell}[Grad_{\ell}F]+ O(a^{2}_{1}) ,\label{grad1}\\
  && (Grad_{\ell}F)_{k}=F'_{k}+(k+1)F_{k+1} ,\quad F'_{k}:=\frac{\partial}{\partial
  \zeta}F_{k}(\zeta) ,\label{grad2}
\end{eqnarray}
and finally the Laplacian map
\begin{eqnarray}
  && \Box \Psi^{\ell}[F]=\Psi^{\ell}[Lap_{\ell}F]+O(a^{2}_{1},c^{2}_{2}) , \label{lm1}\\
  &&(Lap_{\ell}F)_{k}=(\zeta^{2}-1)F''_{k}+(d+1+4k)\zeta
  F'_{k}+[\ell+k(d+2\ell-k)]F_{k}\nonumber\\&&+2\zeta(k+1)^{2}
  F_{k+1}+2(\ell-k+1)F'_{k-1},\label{lm2}\\
  &&\Box F_{k}(\zeta)=(\zeta^{2}-1)F''_{k}+(d+1)\zeta F'_{k}.\quad\label{lm3}
\end{eqnarray}
For the derivation of these maps we used several relations for the
derivatives of the bitensors listed in the Appendix A.

Now we start to explore the restrictions on the sets of functions
$F$, $G$ and $H$ in (\ref{Kk}) given by the following equations for
the propagator in de Donder gauge
\begin{eqnarray}
  && \Box_{a}\Box_{a}K^{\ell}(a,c;\zeta)=0 ,\label{dtr} \\
  && \nabla^{\mu}\frac{\partial}{\partial a^{\mu}}K^{\ell}(a,c;\zeta)
  =\frac{1}{2}(a\nabla)\Box_{a}K^{\ell}(a,c;\zeta) ,\label{gcond} \\
  && \left(\Box+\ell\right)
  K^{\ell}(a,c;\zeta)-a^{2}\Box_{a}K^{\ell}(a,c;\zeta)
  =\Delta_{\ell}(\Delta_{\ell}-d)K^{\ell}(a,c;\zeta)\label{eqofmo} , \,\zeta\neq 1.\label{eqofm}
\end{eqnarray}
Note that we do not write the delta function with the corresponding
projector in the right hand side of (\ref{eqofm}) assuming that we
will focus on the  solution of the equation of motions with the
right normalized delta function singularity at $\zeta\rightarrow 1$.
Actually we have to watch the tensorial structure of this
singularity also. Taking into account that the leading term of the
projector always is
$\delta(z_{1},z_{2})(a^{\mu}g_{\mu\nu}(z_{1})c^{\nu})^{\ell}$ we see
that the most important and singular function in our propagator is
the function $F_{0}$ from $\Psi^{\ell}[F]=I^{\ell}_{1}F_{0}+\dots$
because $I_{1}(a,b;\zeta)\rightarrow
-a^{\mu}g_{\mu\nu}(z_{1})c^{\nu}$ when $\zeta \rightarrow 1$ and
$\ell$ is even. The structure of the other terms in (\ref{eqofm})
will be fixed automatically after considering the two other
conditions (\ref{dtr}) and (\ref{gcond}). Substituting (\ref{Kk}) in
the double tracelessness condition (\ref{dtr}) using recursively the
trace map (\ref{trace}) and neglecting all trace terms, we obtain
the following relations
\begin{eqnarray}
  && \left[\Box_{a}\right]^{2}K^{\ell}=I_{2c}^{4}\Psi^{\ell-4}
  \left[\Theta\right]+O(a^{2}_{1}, c^{2}_{2})=0 , \\
   &&\Theta_{k}= (Tr_{\ell-2}Tr_{\ell}F)_{k}+
   4(\alpha_{0}-2)(Tr_{\ell-2}G)_{k}+8(\alpha_{0}-2)(\alpha_{0}-4)H_{k} , \\
  &&
  (Tr_{\ell-2}Tr_{\ell}F)_{k}=(\ell-k-2)(\ell-k-3)(Tr_{\ell}F)_{k}\nonumber\\&&
  +2(k+1)(\ell-k-3)\zeta(Tr_{\ell}F)_{k+1}+(k+2)(k+1)(\zeta^{2}-1)(Tr_{\ell}F)_{k+2}
  .\,
\end{eqnarray}
We see that from the conditions $\Theta_{k}(\zeta)=0$ we can express
$H_{k}(\zeta)$ as functions of the $G_{k}(\zeta)$ and $F_{k}(\zeta)$
. The next neglected term of order $O(c^{2}_{2})$ will express the
functions $T_{k}$ from the expansion (\ref{Kk}) in a similar way. It
is clear that due to the double tracelessness condition we have only
two free sets of functions $F_{k}(\zeta)$ and $G_{k}(\zeta)$.

Then we consider the de Donder gauge condition. After insertion of
(\ref{Kk}) into the Eq. (\ref{gcond}) and using (\ref{div}) and
(\ref{trace}) we obtain
\begin{eqnarray}
  && \nabla^{\mu}\frac{\partial}{\partial
  a^{\mu}}K^{\ell}-\frac{1}{2}(a\nabla)\Box_{a}K^{\ell}\nonumber\\
  &&=I_{2c}\Psi^{\ell-1}[M_{k}]-\frac{1}{2}(a\nabla)I^{2}_{2c}\Psi^{\ell-2}
  [N_{k}]+O(a^{2}_{1}, c^{2}_{2}) , \\
&& M_{k}(\zeta)=(Div_{\ell}F)_{k}+2(k+2)G_{k}+2G'_{k-1} ,\\
&& N_{k}(\zeta)=(Tr_{\ell}F)_{k}+2\alpha_{0}G_{k} .
\end{eqnarray}
Using the gradient map (\ref{grad1}),(\ref{grad2}) and the formulas
from the Appendix A we can derive the following relation
\begin{equation}\label{grad3}
    (a\nabla)I^{2}_{2c}\Psi^{\ell-2}[N_{k}]=I_{2c}\Psi^{\ell-1}
    [(k+2)N_{k}+N'_{k-1}]+O(a^{2}_{1}) .
\end{equation}
This leads to the final equation
\begin{eqnarray}
  && (Div_{\ell}F)_{k}
  -\frac{1}{2}\left[(k+2)(Tr_{\ell}F)_{k}+(Tr_{\ell}F)'_{k-1}\right]\nonumber\\
  &&=(\alpha_{0}-2)\left[(k+2)G_{k}+G'_{k-1}\right] .
\end{eqnarray}
From the latter we can express all $G_{k}$ as the functions of the
unconstrained $F_{k}$.

So we understood that double tracelessness and de Donder condition
fix the propagator and we have freedom only in the first set of
functions $F_{k}$ forming the leading term $\Psi^{\ell}[F]$ of the
propagator $K^{\ell}(a,b;\zeta)$ (\ref{Kk}). This last free set we
can fix only from the dynamical equation of motion (\ref{eqofmo})
using the Laplacian map (\ref{lm1})-(\ref{lm3})
\begin{eqnarray}
  && (Lap_{\ell}F)_{k}+\ell F_{k}-\Delta_{\ell}(\Delta_{\ell}-d)F_{k}=0 \Rightarrow\\
  &&(\zeta^{2}-1)F''_{k}+(d+1+4k)\zeta F'_{k}+[2\ell+k(d+2\ell-k)]F_{k}\nonumber\\
  &&+2\zeta(k+1)^{2}F_{k+1}+2(\ell-k+1)F'_{k-1}=\Delta_{\ell}(\Delta_{\ell}-d)F_{k}\label{fk}
  .
\end{eqnarray}
These equations are again recursive and will express the higher
functions $F_{k}$ through the lower ones.

As an initial condition we use that $F_{0}$ satisfies the wave
equation of a scalar field of dimension $\Delta_{\ell}$
\begin{eqnarray}
  && (\zeta^{2}-1)F''_{0}+(d+1)\zeta F'_{0}-
  \Delta_{\ell}(\Delta_{\ell}-d)F_{0}=0 , \quad\zeta\ne 1 .\label{f0}
\end{eqnarray}
Then from (\ref{fk}) with $k=0$ follows
\begin{equation}\label{f1}
    F_{1}(\zeta)=-\frac{\ell}{\zeta}F_{0} .
\end{equation}
This result can be generalized to an ansatz of a "main" term and a
"small" (at $\zeta\rightarrow \infty$) term
\begin{eqnarray}
  && F_{k}=c_{k}\zeta^{-k}F_{0}+f_{k} , \label{fmain}\\
  && c_{k}=(-1)^{k}\binom{\ell}{k} , \quad f_{0}=f_{1}=0 .
\end{eqnarray}
We introduce the differential operator $D_{k}$ to abbreviate
(\ref{fk})
\begin{eqnarray}
  && D_{k}(F_{k})+2\zeta(k+1)^{2}F_{k+1}+2(\ell-k+1)F'_{k-1}=0 \label{neq}
\end{eqnarray}
Inserting only the main part of (\ref{fmain}) into (\ref{neq}) we
obtain a residual expression
\begin{eqnarray}
  &&
  Res(F_{k})=\frac{kc_{k}}{\zeta^{k+2}}\left(2\zeta
  F'_{0}-(k+1)F_{0}\right) .
\end{eqnarray}
It is smaller by $O(\zeta^{-2})$ at $\zeta\rightarrow \infty$ than
the main term of $F_{k}$ and arises in
\begin{eqnarray}
  && (\zeta^{2}-1)F''_{k}=c_{k}\zeta^{-k}\left\{(\zeta^{2}-1)F''_{0}
  -2k\zeta^{-1}(\zeta^{2}-\underline{1})F'_{0}\right.\nonumber\\
  &&\left.\qquad\qquad\qquad\qquad\qquad\qquad+k(k+1)\zeta^{-2}(\zeta^{2}-\underline{1})F_{0}\right\}
\end{eqnarray}
from the two underlined terms. Thus we end up with
\begin{eqnarray}
  && D_{k}(f_{k})+2\zeta(k+1)^{2}f_{k+1}+2(\ell-k+1)f'_{k-1}=-Res(F_{k})
\end{eqnarray}
which can be rewritten as
\begin{eqnarray}
  &&
  f_{k+1}=\frac{1}{2(k+1)^{2}\zeta}\left\{-Res(F_{k})-D_{k}(f_{k})-2(\ell-k+1)f'_{k-1}\right\}
  .
\end{eqnarray}
The first cases are
\begin{eqnarray}
  && f_{0}=f_{1}=0 , \\
  && f_{2}=\frac{\ell}{4\zeta^{4}}\left(\zeta F'_{0}-F_{0}\right) ,\\
  &&
  f_{3}=-\frac{1}{18\zeta}\left[\frac{\ell(\ell-1)}{\zeta^{4}}\left(2\zeta
  F'_{0}-3F_{0}\right)+D_{2}(f_{2})\right] .
\end{eqnarray}
The main term of (\ref{fmain}) can be summed and gives for the
propagator without trace terms
\begin{eqnarray}
  && \left(I_{1}-\frac{1}{\zeta}I_{2}\right)^{\ell}F_{0}(\zeta)\label{mainterm}
\end{eqnarray}

In the bulk-to-boundary limit (see the next section for details)
this reduces to Dobrev's propagator \cite{Dobrev} (without trace
terms).

 The solution for the wave equation (\ref{f0}) is \cite{Freed,RM3}
\begin{equation}\label{fm}
F_{0}=C\zeta^{-\Delta_{\ell}}{}_{2}F_{1}(\frac{1}{2}\Delta_{\ell},\frac{1}{2}
(\Delta_{\ell}+1);\Delta_{\ell}-\mu+1;\zeta^{-2}) .
\end{equation}
We set the normalization constant $C$ equal to
\begin{equation}\label{C}
    C=\frac{\Gamma(\frac{1}{2}\Delta_{\ell})\Gamma(\frac{1}{2}(\Delta_{\ell}+1))}
    {(4\pi)^{\mu+\frac{1}{2}}\Gamma(\Delta_{\ell}-\mu+1)} .
\end{equation}
 and use the fact that for $\zeta\rightarrow 1$
\begin{eqnarray}
  &&{}_{2}F_{1}(\frac{1}{2}\Delta_{\ell},\frac{1}{2}(\Delta_{\ell}+1);\Delta_{\ell}-\mu+1;\zeta^{-2})\nonumber \\
  &&= \frac{\Gamma(\Delta_{\ell}-\mu+1)\Gamma(\mu-\frac{1}{2})}{\Gamma(\frac{1}{2}\Delta_{\ell})
  \Gamma(\frac{1}{2}(\Delta_{\ell}+1))}(\zeta^{2}-1)^{-\mu+\frac{1}{2}}+O(1)
   , \quad \Re\mu>\frac{1}{2} , \label{glim}
\end{eqnarray}
and
\begin{equation}
\Box\frac{\Gamma(\mu-\frac{1}{2})}{(4\pi)^{\mu+\frac{1}{2}}}(\zeta^{2}-1)^{-\mu+\frac{1}{2}}
  =-\delta(z_{1},z_{2})+\text{regular terms} ,\label{delta}
  \end{equation}

So we  prove that $F_{0}(\zeta)$ appears as the kernel for the
inverse wave operator $(-\Box + m^{2})$ for the massive scalar field
in Euclidian $AdS_{d+1}$ space with
$m^{2}=\Delta_{\ell}(\Delta_{\ell}-d)$.

\section{Bulk-to-boundary limit}

Now we can take the boundary limit and obtain the spin $\ell$
bulk-to-boundary propagator from the bulk-to-bulk propagator
directly. For this purpose we mention that the boundary of $AdS$
space is approached in the limit
\begin{equation}\label{lim}
    z^{0}\rightarrow 0 ,
\end{equation}
which is connected with the limit $\zeta\rightarrow\infty$ due to
\begin{equation}\label{zlim}
    \lim_{z^{0}_{2}\rightarrow 0}
    2z^{0}_{1}z^{0}_{2}\zeta(z_{1},z_{2})=(z^{0}_{1})^{2}+(\vec{z}_{1}-\vec{z}_{2})^{2}
    .
\end{equation}
Then following the explanation of the previous section we see that
at the boundary only the main term (\ref{mainterm}) survives and we
get
\begin{eqnarray}
  &&\lim_{\substack{z^{0}_{2}\rightarrow
0  \\ c_{\mu}=(0,\vec{c})} }(z^{0}_{2})^{\ell-\Delta}
  \left(I_{1}-\frac{1}{\zeta}I_{2}\right)^{\ell}F_{0}(\zeta)
  =2^{\Delta}C\frac{(z^{0}_{1})^{d-2}}{\left[(z^{0}_{1})^{2}
  +(\vec{z}_{1}-\vec{z}_{2})^{2}\right]^{\Delta}}
  \left[R(a,\vec{c};z_{1}-\vec{z_{2}})\right]^{\ell} ,\qquad\label{Flim}\\
  &&R(a,\vec{c};z_{1}-\vec{z_{2}})=<\vec{a},\vec{c}>-2\frac{(a,z_{1})<\vec{z}_{1}-\vec{z}_{2},\vec{c}>}
  {(z^{0}_{1})^{2}+(\vec{z}_{1}-\vec{z}_{2})^{2}} .
\end{eqnarray}
Here we introduced the $d+1$ and $d$ dimensional Euclidian scalar
products
\begin{equation}\label{sp}
    (a,z)=\sum^{d}_{\mu=0}a^{\mu}z_{\mu} , \quad
    <\vec{c},\vec{z}>=\sum^{d}_{i=1}c^{i}z_{i}
\end{equation}
and the Jacobian tensor
\begin{equation}\label{jt}
    R_{\mu\nu}(z)=\delta_{\mu\nu}-2\frac{z_{\mu}z_{\nu}}{(z,z)} .
\end{equation}
We see that the limit (\ref{Flim}) really produces
 Dobrev's \cite{Dobrev} boundary-to-bulk propagator
without trace terms.

Actually we need only this leading term because all other trace
terms  depend on the  gauge condition (\ref{gcond}) applied to the
bulk dependent side of the right hand side of (\ref{Flim}). On the
other hand we can fix the trace terms by requiring the tensor fields
to approach a certain tensor type on the boundary. In the case of
irreducible $d$ dimensional $CFT$ currents we have to claim
tracelessness with respect to the indices contracted with $\vec{c}$
\begin{eqnarray}
  && \Box_{\vec{c}}G^{(\ell)}_{AdS/CFT}(a,\vec{c};z)=\frac{\partial^{2}}{\partial
    \vec{c}\partial\vec{c}}\left(G^{(\ell)}_{m}(a,\vec{c};z)+\textnormal{trace
    terms}\right)=0 ,\label{adscft} \\
  &&G^{(\ell)}_{m}(a,\vec{c};z)=\frac{(z^{0})^{d-2}}{(z,z)^{\Delta}}
  \left[R(a,\vec{c};z)\right]^{\ell} .
\end{eqnarray}
Here we omit the normalization factor $2^{\Delta}C$ and put for
simplicity $\vec{z_{2}}=0$ and $z^{\mu}_{1}=z^{\mu}$ (we can always
restore the right dependence on the boundary coordinate
$\vec{z_{2}}$ using translation invariance in the flat boundary
space).

Then considering the boundary limit of the $I_{3}$ and $I_{4}$
dependent terms we can easily render the propagator (\ref{adscft})
traceless on the boundary by the projection\footnote{In this section
we used the exact expression for Christoffel symbols\\
$\Gamma^{\lambda}_{\mu\nu}=\frac{1}{z^{0}}\left(\delta^{\lambda}_{0}
\delta_{\mu\nu}-2\delta^{0}_{(\mu}\delta^{\lambda}_{\nu)}\right)$
and the $AdS$ trace rule
$\Box_{a}=(z^{0})^{2}\delta^{\mu\nu}\frac{\partial^{2}}{\partial
a^{\mu}\partial a^{\nu}}$ }
\begin{eqnarray}
  &&
  G^{(\ell)}_{AdS/CFT}(a,\vec{c};z)=G^{(\ell)}_{m}(a,\vec{c};z)-\frac{(a,a)-[R^{0}(a;z)]^{2}}
  {2(\alpha_{0}-1)(z^{0})^{2}}\Box_{a}G^{(\ell)}_{m}(a,\vec{c};z)\nonumber\\
&&\hspace{8cm}+ O(a^{4})+O(c^{4}) .\label{trads}
\end{eqnarray}
The complete polynomial expression for
$G^{(\ell)}_{AdS/CFT}(a,\vec{c};z)$ is presented in the Appendix B,
Eqn. (\ref{b10}). But here we consider only the first order trace
term
\begin{eqnarray}
 &&\Box_{a}G^{(\ell)}_{m}(a,\vec{c};z)=\ell(\ell-1)\frac{(z^{0})^{d}}{(z,z)^{\Delta}}<\vec{c},\vec{c}>
  \left[R(a,\vec{c};z)\right]^{\ell-2} ,\\
  &&R^{0}(a;z)=a^{\mu}R_{\mu}^{0}(z)=a^{0}-2\frac{z^{0}(a,z)}{(z,z)}
  , \quad \alpha_{0}=d+2\ell-3 .
\end{eqnarray}
The important point of this consideration is the following: The
expression (\ref{trads}) is automatically traceless on the $AdS$
side.
\begin{equation}\label{adstr}
    \Box_{a}G^{(\ell)}_{AdS/CFT}(a,\vec{c};z)=0 ,
\end{equation}
due to the relations
\begin{equation}\label{rel}
\delta^{\mu\nu}R_{\mu}^{0}(z)R_{\nu}^{0}(z)=1 \quad , \quad
\delta^{\mu\nu}R_{\mu}^{0}(z)R_{\nu}(\vec{c};z)=0
\end{equation}
This is natural because the original   bulk-to-bulk basis
$\left\{I_{i}(a,c;\zeta)\right\}_{i=1}^{4}$ was symmetric with
respect to the $a\leftrightarrow c$ exchange . Then we see that this
projection in agreement with de Donder gauge condition (\ref{gcond})
(for traceless case) leads to the transverse-traceless
bulk-to-boundary propagator (\ref{trads}) for all higher spin fields
on $AdS$ side. For proving this we have to calculate several
relations in first order of $(a,a)$ and $<\vec{c},\vec{c}>$ (see
details in Appendix B)
\begin{eqnarray}
&& \nabla^{\mu}\frac{\partial}{\partial
a^{\mu}}G^{(\ell)}_{m}(a,\vec{c};z)=
  \ell(\ell-1)\frac{(z^{0})^{d-1}}{(z,z)^{\Delta}}<\vec{c},\vec{c}>
  \left[R(a,\vec{c};z)\right]^{\ell-2}R^{0}(a;z) ,\label{tr1}\\
  && a^{\mu}\nabla_{\mu}\Box_{a}G^{(\ell)}_{m}(a,\vec{c};z)
  =\ell(\ell-1)\frac{(z^{0})^{d-1}}{(z,z)^{\Delta}}<\vec{c},\vec{c}>
  \left[R(a,\vec{c};z)\right]^{\ell-2}R^{0}(a;z)(\alpha_{0}-1) ,\qquad \label{tr2}\\
&&\nabla^{\mu}\frac{\partial}{\partial
a^{\mu}}\frac{[R^{0}(a;z)]^{2}}
  {(z^{0})^{2}}\Box_{a}G^{(\ell)}_{m}(a,\vec{c};z)= 0 . \label{tr3}
\end{eqnarray}
Putting all together we obtain
\begin{equation}\label{final}
    \nabla^{\mu}\frac{\partial}{\partial
a^{\mu}}G^{(\ell)}_{AdS/CFT}(a,\vec{c};z)= 0 .
\end{equation}
So we see that the  Goldstone  mode (non transverseness of the
propagator) can not be visible after trace projection on the
boundary side corresponding to the case of the traceless currents in
the large $N$ limit of the $O(N)$ sigma model.

The next interesting question which we can ask is the transversal
property of the bulk-to-boundary propagator on the boundary side.
The answer is negative. The divergence on the boundary side of the
traceless bulk-to-boundary propagator is not zero  and equals  a
gauge term (gradient) with respect to the bulk gauge invariance.
Using the formulas from  Appendix B one can check that
\begin{eqnarray}
  && \frac{\partial}{\partial \vec{z}}\cdot \frac{\partial}{\partial \vec{c}}
  G^{(\ell)}_{AdS/CFT}(a,\vec{c};z)
  =a^{\mu}\nabla_{\mu}
  \Lambda^{(\ell-1)}(a,\vec{c};z)
  ,\\
  &&\Lambda^{(\ell-1)}(a,\vec{c};z)=2\ell\frac{(\alpha_{0}-1)(\ell+d-1)
  -2(\ell-1)}{\alpha_{0}^{2}-1}\frac{(z^{0})^d}{(z,z)^{\Delta_{\ell}+1}}
  [R(a,\vec{c};z)]^{\ell-1} .\qquad
\end{eqnarray}
We see that the boundary trace projection generates the bulk gauge
term on the boundary side and is equivalent to the residual on-shell
gauge fixing preserving the bulk side de Donder off-shell gauge
(this property of the bulk-to-boundary propagator was mentioned in
\cite{Mikh} and in \cite{fr} for the vector field case).

Finalizing our consideration we can define now the $CFT$ propagator
from (\ref{trads}) by $a^{0}=0$ and the limit $z^{0}\rightarrow 0$.
Due to the vanishing of $R^{0}(a;z)$  in this limit we get
\begin{eqnarray}
  && G^{(\ell)}_{CFT}(\vec{a},\vec{c};\vec{z})=\lim_{z^{0}\rightarrow 0}
  (z^{0})^{2-d}G^{(\ell)}_{AdS/CFT}(\vec{a},\vec{c};z) \nonumber\\
  &&= G^{(\ell)}_{m}(\vec{a},\vec{c};\vec{z})-\frac{<\vec{a},\vec{a}>}
  {2(\alpha_{0}-1)}\Box_{\vec{a}}G^{(\ell)}_{m}(\vec{a},\vec{c};
  \vec{z})+O(<\vec{a},\vec{a}>^2)\label{fincft} \\
\end{eqnarray}
Thus the limit (\ref{fincft}) defines the correct $CFT$ two point
function for traceless conserved\footnote{Note that the gradient of
the gauge term also vanishes on the boundary because
$a^{\mu}\nabla_{\mu}\Lambda^{(\ell-1)}(a,\vec{c};z) \sim
R^{0}(a;z)$.} currents.

So we prove that the boundary limit of our bulk-to-bulk propagator
in the de Donder gauge is in agreement with the bulk-to-boundary
propagator obtained from the $AdS$ isometry group representation
theory.

\subsection*{Acknowledgements}
\quad This work is supported in part by the German
Volkswagenstiftung. The work of R.~M. was supported by DFG (Deutsche
Forschungsgemeinschaft) and in part by the INTAS grant \#03-51-6346.

\section*{Appendix A}
\setcounter{equation}{0}
\renewcommand{\theequation}{A.\arabic{equation}}

In this article we use the following rules and relations for
$\zeta(z,z')$, $I_{1a}$, $I_{2c}$ and the bitensorial basis
$\{I_{i}\}^{4}_{i=1}$
\begin{eqnarray}
  && \Box\zeta=(d+1)\zeta ,\quad \nabla_{\mu}\partial_{\nu}\zeta=g_{\mu\nu}\zeta ,
  \quad g^{\mu\nu}\partial_{\mu}\zeta\partial_{\nu}\zeta=\zeta^{2}-1 ,\\
  &&   \partial_{\mu}\partial_{\nu'}\zeta
  \nabla^{\mu}\zeta=\zeta\partial_{\nu'}\zeta ,\quad
  \partial_{\mu}\partial_{\nu'}\zeta \nabla^{\mu}\partial_{\mu'}\zeta
  =g_{\mu'`\nu'}+\partial_{\mu'}\zeta\partial_{\nu'}\zeta ,\\
&&\nabla_{\mu}\partial_{\nu}\partial_{\nu'}\zeta \nabla^{\mu}\zeta
  =\partial_{\nu}\zeta\partial_{\nu'}\zeta ,\quad
  \nabla_{\mu}\partial_{\nu}\partial_{\nu'}\zeta
  =g_{\mu\nu}\partial_{\nu'}\zeta ,\\
&&\frac{\partial}{\partial a^{\mu}}I_{1a}\frac{\partial}{\partial
a_{\mu}}I_{1a}=\zeta^{2}-1 ,\quad \frac{\partial}{\partial
a^{\mu}}I_{1}\frac{\partial}{\partial
a_{\mu}}I_{1a}=\zeta I_{2c} ,\\
&&\frac{\partial}{\partial a^{\mu}}I_{1}\frac{\partial}{\partial
a_{\mu}}I_{1}=c^{2}_{2}+ I_{2c}^{2} , \, \frac{\partial}{\partial
a^{\mu}}I_{1}\frac{\partial}{\partial
a_{\mu}}I_{2}=\zeta I_{2c}^{2} ,\,\Box_{a}I_{4}=2(d+1)c^{2}_{2} ,\\
&&\frac{\partial}{\partial a^{\mu}}I_{2}\frac{\partial}{\partial
a_{\mu}}I_{2}=(\zeta^{2}-1)I_{2c}^{2} ,\quad
\Box_{a}I_{3}=2(d+1)I_{2c}^{2}+2c^{2}_{2}(\zeta^{2}-1) ,\\
&&\nabla^{\mu}\frac{\partial}{\partial a^{\mu}}I_{1}=(d+1)I_{2c}
,\,\nabla^{\mu}\frac{\partial}{\partial a^{\mu}}I_{2}=(d+2)\zeta
I_{2c},\quad\nabla^{\mu} I_{1}\partial_{\mu}
\zeta=I_{2} ,\\
&&\nabla^{\mu}\frac{\partial}{\partial
a^{\mu}}I_{3}=4I_{1}I_{2c}+2(d+2)\zeta c^{2}_{2}I_{1a}
,\quad\nabla^{\mu} I_{2}\partial_{\mu} \zeta=2\zeta I_{2} ,\\
&&\frac{\partial}{\partial a_{\mu}} I_{1}\partial_{\mu} \zeta=\zeta
I_{2c} ,\quad \frac{\partial}{\partial a_{\mu}} I_{2}\partial_{\mu}
\zeta=(\zeta^2-1) I_{2c} ,\,\frac{\partial}{\partial a_{\mu}}
I_{1}\nabla_{\mu} I_{1}=I_{1} I_{2c} ,\,\,\,\,\,\\
&&\frac{\partial}{\partial a_{\mu}} I_{1}\nabla_{\mu}
I_{2}=I_{2c}\left(\zeta I_{1}+I_{2}\right)+c^{2}_{2}I_{1a}
,\frac{\partial}{\partial a_{\mu}} I_{2}\nabla_{\mu}
I_{1}=I_{2c}I_{2} ,\\
&&\frac{\partial}{\partial a_{\mu}} I_{2}\nabla_{\mu} I_{2}=2\zeta
I_{2c}I_{2} ,\quad \nabla^{\mu} I_{1}\nabla_{\mu}
I_{1}=a^{2}_{1}I_{2c} ,\quad \Box I_{1}=I_{1} ,\\
&&\nabla^{\mu} I_{1}\nabla_{\mu} I_{2}=I_{2}I_{1}+ a^{2}_{1}\zeta
I_{2c} ,\quad \Box I_{2}=(d+2)I_{2}+2\zeta
I_{1} ,\quad\\
&&\nabla^{\mu} I_{2}\nabla_{\mu} I_{2}=I_{2}^{2}+2\zeta
I_{1}I_{2}+a^{2}_{1}I_{2c}^{2}\zeta^{2}+c^{2}_{2}I_{1a}^{2}
,\quad\nabla^{\mu} I_{2}\partial_{\mu}\zeta=2\zeta I_{2} ,\\
&&a^{\mu}\nabla_{\mu}I_{1a}=a^{2}\zeta ,\quad
a^{\mu}\nabla_{\mu}I_{2c}=I_{1},\quad
a^{\mu}\nabla_{\mu}I_{1}=a^{2}I_{2c}
,\\&&a^{\mu}\nabla_{\mu}I_{2}=a^{2}\zeta I_{2c}+I_{1a}I_{1}
,\quad\nabla^{\mu} I_{1}\partial_{\mu}\zeta=I_{2}.
\end{eqnarray}
\section*{Appendix B}
\setcounter{equation}{0}
\renewcommand{\theequation}{B.\arabic{equation}}
Here we prove the relations (\ref{tr1})-(\ref{final}). The more
transparent way of working with the boundary-to-bulk propagator for
higher spins is to introduce two additional objects
\begin{eqnarray}
  && \phi^{0}(z)=\frac{z^{0}}{(z,z)} ,\label{b1}\\
  && \psi(\vec{c},z)=\frac{<\vec{c},\vec{z}>}{(z,z)} ,\label{b2}
\end{eqnarray}
satisfying the following relations
\begin{eqnarray}
  && a^{\mu}\partial_{\mu}\phi^{0}(z)=\frac{R^{0}(a;z)}{(z,z)} ,\quad
    a^{\mu}\partial_{\mu}\psi(\vec{c},z)=\frac{R(a,\vec{c};z)}{(z,z)}  ,\label{b3}\\
  && \Box \phi^{0}(z)=-(d-1)\phi^{0}(z) , \quad \Box \psi(\vec{c},z)=0 , \label{b4}\\
  &&a^{\mu}a^{\nu}\nabla_{\mu}\partial_{\nu}\psi(\vec{c},z)=
   2[\phi^{0}(z)]^{-1}a^{\mu}\partial_{\mu}\phi^{0}(z)a^{\nu}\partial_{\nu}\psi(\vec{c},z) ,\label{b5}\\
  && \nabla^{\mu}\phi^{0}(z)\partial_{\mu}\phi^{0}(z)=(\phi^{0})^{2}
  ,\quad
  \nabla^{\mu}\phi^{0}(z)\partial_{\mu}\psi(\vec{c},z)=0\label{b6}\\
  &&
  \nabla^{\mu}\psi(\vec{c},z)\partial_{\mu}\psi(\vec{c},z)=(\phi^{0})^{2}<\vec{c},\vec{c}>
  , \label{b7}\\&&\quad \Box=\nabla^{\mu}\partial_{\mu},\quad \nabla^{\mu}\left\{\substack{\phi^{0}(z)\\
  \psi(\vec{c},z)}
  \right\}=g^{\mu\nu}\partial_{\nu}\left\{\substack{\phi^{0}(z)\\ \psi(\vec{c},z)}
  \right\} ,\label{b8}\\
  &&\nabla_{\mu}\partial_{\nu}=
  g^{\mu\nu}\left(\partial_{\mu}\delta^{\lambda}_{\nu}-
  \Gamma^{\lambda}_{\mu\nu}\right)\partial_{\lambda} ,\quad
  \Gamma^{\lambda}_{\mu\nu}=\frac{1}{z^{0}}\left(\delta^{\lambda}_{0}\delta_{\mu\nu}
  -\delta^{\lambda}_{\mu}\delta_{\nu 0}-\delta^{\lambda}_{\nu}\delta_{\mu
  0}\right).
\end{eqnarray}
Then using (\ref{b1})-(\ref{b3}) we can rewrite the $AdS/CFT$
bulk-to-boundary propagator (\ref{trads}) in the following complete
form
\begin{eqnarray}
  && G^{(\ell)}_{AdS/CFT}(a,\vec{c};z)=(\phi^{0}(z))^{d-2}
  \sum^{[\ell/2]}_{k=0}\frac{(-\ell)_{2k}}{2^{2k}k!(\frac{1-\alpha_{0}}{2})_{k}}
  [a^{\mu}\partial_{\mu}\psi(\vec{c},z)]^{\ell-2k}
\nonumber\\
&&\times
\left[<\vec{c},\vec{c}>\left(a^{\mu}a_{\mu}(\phi^{0}(z))^{2}-[a^{\mu}
\partial_{\mu}\phi^{0}(z)]^{2}\right)\right]^{k}
  .\qquad\quad\label{b10}
\end{eqnarray}
After that the proof of the condition
\begin{equation}
\nabla^{\mu}\frac{\partial}{\partial
a^{\mu}}G^{(\ell)}_{AdS/CFT}(a,\vec{c};z)=0
\end{equation}
reduces to the differentiation of the right hand side of with the
covariant Leibniz rules and use of the relations
(\ref{b4})-(\ref{b6}).

For taking the divergence on the boundary side of (\ref{trads}) or
(\ref{b10}) we need the following identities for $\psi(\vec{c},z)$
and $\phi^{0}(z)$
\begin{eqnarray}
  && \vec{\psi}(z)=\frac{\partial}{\partial \vec{c}}\psi(\vec{c},z)=\frac{\vec{z}}{(z,z)}
   ,\quad\vec{\psi}(z)\cdot\vec{\psi}(z)=\frac{1}{(z,z)}-[\phi^{0}(z)]^{2} ,\label{b13}\\
   &&\frac{\partial}{\partial \vec{z}}\phi^{0}(z)=-2\phi^{0}(z)\vec{\psi}(z) ,
   \quad a^{\mu}\partial_{\mu}\frac{\partial}{\partial \vec{z}}\cdot\vec{\psi}(z)
   =a^{\mu}\partial_{\mu}\frac{d-2}{(z,z)}+4\phi^{0}(z)a^{\mu}\partial_{\mu}\phi^{0}(z)
   ,
   \quad\qquad\\
  &&\frac{\partial}{\partial \vec{z}}\phi^{0}(z)\cdot a^{\mu}\partial_{\mu}\vec{\psi}(z)
  =2[\phi^{0}(z)]^{2}a^{\mu}\partial_{\mu}\phi^{0}(z)-\phi^{0}(z)a^{\mu}\partial_{\mu}
  \frac{1}{(z,z)} , \\
&& a^{\mu}\partial_{\mu}\frac{\partial}{\partial \vec{z}}
\psi(\vec{c},z)\cdot a^{\nu}\partial_{\nu}\vec{\psi}(z)= 2
\phi^{0}(z)a^{\mu}\partial_{\mu}\phi^{0}(z)a^{\mu}\partial_{\mu}\psi(\vec{c},z)\nonumber\\
&&\hspace{5cm}-2\psi(\vec{c},z)
\left([\phi^{0}(z)]^{2}a^{\mu}a_{\mu}-[a^{\mu}\partial_{\mu}\phi^{0}(z)]^2\right)\label{b14}
\end{eqnarray}
Then performing boundary differentiation of (\ref{b10}) and using
(\ref{b13})-(\ref{b14}) we obtain
\begin{eqnarray}
  && \frac{\partial}{\partial \vec{z}}\cdot \frac{\partial}{\partial \vec{c}}G^{(\ell)}_{AdS/CFT}(a,\vec{c};z)
  =a^{\mu}\nabla_{\mu}
  \Lambda^{(\ell-1)}(a,\vec{c};z) + O(<\vec{c},\vec{c}>)
  ,\\
  &&\Lambda^{(\ell-1)}(a,\vec{c};z)=2\ell\frac{(\alpha_{0}-1)
  (\ell+d-1)-2(\ell-1)}{\alpha_{0}^{2}-1}[\phi^{0}(z)]^d
  [a^{\mu}\partial_{\mu}\psi(\vec{c},z)]^{\ell-1} .\quad\qquad
\end{eqnarray}

\end{document}